\documentstyle{mn}
\input{psfig}
\newcommand{\etal}{{et al.~}}

\newcommand{\kmsmpc}{\>{\rm km}\,{\rm s}^{-1}\,{\rm Mpc}^{-1}}
\newcommand{\kms}{\>{\rm km}\,{\rm s}^{-1}}

\newcommand{\Msun}{\>{\rm M_{\odot}}}

\newcommand{\beq}{\begin{equation}}
\newcommand{\eeq}{\end{equation}}

\newcommand{\rp}{\,R_{\rm proj}/R_{180}}
\newcommand{\rproj}{\,R_{\rm proj}/R_{180}}
\newcommand{\col}{\,^{0.1}(g-r)}
\newcommand{\calC}{{\cal C}}

\def\gtsima{$\; \buildrel > \over \sim \;$}
\def\ltsima{$\; \buildrel < \over \sim \;$}
\def\prosima{$\; \buildrel \propto \over \sim \;$}
\def\gsim{\lower.7ex\hbox{\gtsima}}
\def\lsim{\lower.7ex\hbox{\ltsima}}
\def\simgt{\lower.7ex\hbox{\gtsima}}
\def\simlt{\lower.7ex\hbox{\ltsima}}
\def\simpr{\lower.7ex\hbox{\prosima}}
\def\la{\lsim}

\def\lta{\la}

\newcommand{\apj}{ApJ}
\newcommand{\apjs}{ApJS}
\newcommand{\aj}{AJ}
\newcommand{\mnras}{MNRAS}
\newcommand{\aap}{A\&A}
\newcommand{\aaps}{A\&AS}

\newdimen\hssize
\newdimen\hdsize
\begin{document}


\title[Satellite Ecology: The Dearth of Environment Dependence]
      {Satellite Ecology: The Dearth of Environment Dependence}
\author[van den Bosch et al.]
       {\parbox[t]{\textwidth}{
        Frank C. van den Bosch$^{1}$\thanks{E-mail: vdbosch@mpia.de}, 
        Anna Pasquali$^{1}$,
        Xiaohu Yang$^{2}$,
        H.J. Mo$^{3}$,\\
        Simone Weinmann$^{4}$,
        Daniel H. McIntosh$^{3}$,
        Daniel Aquino$^{1}$}
        \vspace*{3pt} \\
       $^1$Max-Planck Institute for Astronomy, K\"onigstuhl 17, D-69117
           Heidelberg, Germany\\
       $^2$Shanghai Astronomical Observatory; the Partner Group of MPA,
           Nandan Road 80,  Shanghai 200030, China\\
       $^3$Department of Astronomy, University of Massachusetts,
           Amherst MA 01003-9305, USA\\
       $^4$Max-Planck Institut f\"ur Astrophysik, Karl Schwarzschild
         Str. 1, Postfach 1317, 85741 Garching, Germany}


\date{}

\pagerange{\pageref{firstpage}--\pageref{lastpage}}
\pubyear{2007}

\maketitle

\label{firstpage}


\begin{abstract}
  Using the Sloan Digital Sky  Survey (SDSS) galaxy group catalogue of
  Yang et al.  (2007), we  study the average colour (representing star
  formation history)  and  average   concentration  (representing mass
  assembly history)  of  satellite galaxies as  function of  (i) their
  stellar  mass, (ii) their  group mass, and (iii) their group-centric
  radius.  We  find that the  colours and  concentrations of satellite
  galaxies are (almost)  completely determined by  their stellar mass.
  In particular,  at fixed  stellar   mass,  the average colours   and
  concentrations of satellite galaxies  are independent of either halo
  mass or halo-centric   radius.   We find  clear   evidence for  mass
  segregation of satellite galaxies in haloes of all masses, and argue
  that this explains why satellites  at smaller halo-centric radii are
  somewhat redder  and  somewhat more concentrated.   In addition, the
  weak colour and  concentration  dependence of satellite galaxies  on
  halo mass  is  simply a reflection of   the fact that  more  massive
  haloes host, on average,  more massive satellites.  Combining  these
  results   with  the fact that  satellite   galaxies are, on average,
  redder and somewhat more  concentrated than central galaxies of  the
  same stellar mass, the  following  picture emerges: galaxies  become
  redder and  somewhat more concentrated once  they fall into a bigger
  halo (i.e., once they  become a satellite galaxy).   This is a clear
  manifestation  of  environment   dependence.   However, there is  no
  indication  that  the  magnitude   of  the   transformation (or  its
  timescale) depends on environment;  a galaxy undergoes a  transition
  when  it becomes a  satellite,  but  it does not  matter  whether it
  becomes  a satellite  of a small   (Milky Way sized)  halo, or  of a
  massive  cluster. We discuss   the implication of  this `dearth'  of
  environment  dependence  for the physical processes  responsible for
  transforming satellite galaxies.
\end{abstract}




\section{Introduction}
\label{sec:intro}

In  the current paradigm of   galaxy  formation, it  is believed  that
virtually all galaxies initially form  as disks due  to the cooling of
gas  with non-zero angular momentum in  virialized dark matter haloes.
During their subsequent  evolution, star-forming disk galaxies may  be
transformed  into non    star-forming early-types via   a variety   of
transformation mechanisms.  Two   disk galaxies of roughly  equal mass
may  merge  to produce a  spheroidal  galaxy  (Toomre \&  Toomre 1972;
Negroponte \& White 1983), or a disk  galaxy may be transformed into a
spheroid due to the  cumulative effect of  many high speed (impulsive)
encounters, called harassment  (Farouki  \& Shapiro 1981;  Moore \etal
1996).  When  a  small halo  is accreted  by a   larger halo its  hot,
diffuse  gas may   be  stripped  thus removing   its  fuel for  future
star-formation (Larson, Tinsley  \& Caldwell 1980; Kauffmann, White \&
Guiderdoni 1993; Balogh, Navarro \& Morris 2000).  Following Balogh \&
Morris  (2000) we  refer  to this process,  that  results in a  fairly
gradual  decline   of the    satellite's  star   formation   rate,  as
strangulation.  When  the  external  pressure  is   sufficiently high,
ram-pressure stripping may  also remove the  entire cold gas reservoir
of the satellite  galaxy (Gunn \&  Gott  1972; Quilis, Moore \&  Bower
2000), resulting in an extremely fast quenching of its star formation.
Finally, along its orbit a satellite galaxy is subject to tidal forces
which may cause tidal stripping and heating.

The efficiencies  of all  these various transformation  mechanisms are
expected to be strong functions of environment. Ram-pressure stripping
requires a  dense inter-galactic medium  and is therefore  believed to
occur predominantly  in massive clusters.   The same holds  for galaxy
harassment, which  requires the  presence of many,  relatively massive
satellite galaxies in  order to be efficient.  Galaxy  merging, on the
other hand,  is thought to be  suppressed in massive  clusters, and to
preferentially occur in group-sized  haloes, while tidal stripping and
heating  should play  a role  in all  dark matter  haloes  with little
dependence on  their virial  mass.  Finally, strangulation  can result
from both ram-pressure  and tides, and it is  difficult to predict how
its efficiency would scale with halo mass.  The various transformation
mechanisms  also  have  very  different outcomes:  while  mergers  and
harassment  are thought  to  transform disk  galaxies into  spheroids,
ram-pressure  stripping   and  strangulation  only   affect  the  star
formation  activity   of  the  galaxy  without   changing  its  global
morphology.  Therefore, an assessment of galaxy properties as function
of  their environment  sheds light  on which  of  these transformation
mechanisms operates in what environment and with what efficiency.

Numerous authors have  investigated  the relation  between environment
and morphology (e.g., Hubble 1936; Oemler 1974; Dressler 1980; Postman
\& Geller 1984; Whitmore,  Gilmore \& Jones  1993; Hashimoto \& Oemler
1999; Goto \etal 2003; McIntosh, Rix \& Caldwell  2004; Kuehn \& Ryden
2005;  Blanton \etal 2005b),  between  environment and  star formation
rate (e.g., Hashimoto \etal  1998; Lewis \etal 2002; Dom\'inguez \etal
2002;   G\'omez \etal 2003;  Balogh \etal   2004a;  Tanaka \etal 2004;
Kauffmann  \etal 2004; Kelm, Focardi  \& Sorrentino 2005), and between
environment and colour   (e.g.,Tanaka \etal 2004;  Balogh \etal 2004b;
Hogg \etal 2004;  Weinmann \etal   2006a;  Blanton \&  Berlind  2007).
These   and other studies have   clearly  established that galaxies in
denser environments are  more massive, redder, more concentrated, less
gas-rich and have  an older stellar  population.  Within  the paradigm
outlined  above,  the  transformation     mechanisms  thus   seem   to
preferentially occur in denser environments.  However, a more detailed
interpretation of  these  findings  is complicated   by the fact  that
various  galaxy properties  are strongly correlated   even  at a fixed
environment.  An important  outstanding question,  therefore, is which
relationships with environment are  truly  causal, and which are  just
reflections of  other    more fundamental correlations   that are  not
environment dependent.

With  the advent  of large  redshift surveys,  such as  the Two-Degree
Field  Galaxy Redshift  Survey (2dFGRS;  Colless \etal  2001)  and the
Sloan  Digital Sky  Survey  (SDSS; York  \etal  2000; Stoughton  \etal
2002), several  authors have started to address  these questions using
multi-variate   statistics.   Most   of  these   studies   agree  that
morphological properties are less strongly related to environment than
star-formation  related properties  such as  colour  and emission-line
flux (Kauffmann \etal 2004; Blanton \etal 2005b; Quintero \etal 2006).
This is also in line  with recent results indicating that the relation
between  star  formation  indicators   and  environment  is  at  least
partially  independent  of   galaxy  morphology  (Balogh  \etal  1998;
Koopmann \& Kenney 1998; Poggianti \etal 1999; Christlein \& Zabludoff
2005).

A  physical interpretation   of  all these  results  in  terms  of the
transformation  mechanisms discussed    above  is   hampered   by  two
shortcomings.   First   of   all,  in  most  studies    of environment
dependence, especially those based  on large redshift surveys  such as
the 2dFGRS or the SDSS, one does not split  the galaxy population into
central galaxies (those at  rest at the centers  of their dark  matter
haloes)  and  satellite galaxies  (those  orbiting a  central galaxy).
This is   important    since most of   the   transformation mechanisms
discussed  above   (tidal   stripping  and    heating,  strangulation,
ram-pressure stripping,  and harassment)  mainly operate on  satellite
galaxies.  However, satellite  galaxies only make  up roughly 20 to 40
percent  of the entire  galaxy population (Mandelbaum  \etal 2006; van
den    Bosch \etal 2007a;  Tinker   \etal  2007; Cacciato \etal 2008).
Therefore, unless one  specifically focuses on satellite galaxies, the
environmental   impact  of  the aforementioned   processes  may not be
apparent.

A second problem that hampers a physical interpretation of the results
is  the use  of non-intuitive  environment  indicators.  Most  studies
parameterize `environment' through  the  projected  number density  of
galaxies,  either within some  fixed  metric aperture,  or out  to the
distance  of the $n^{\rm th}$  nearest neighbor (with $n$ typically in
the range 5-10).  Although these environment indicators are relatively
easy to measure from large galaxy redshift surveys, they are difficult
to interpret in terms of more physical  quantities (see discussions in
Kauffmann \etal 2004  and  Weinmann \etal 2006a).  Arguably   the most
natural environment indicators are the halo  mass and the halo-centric
distance, with the virial radius $R_{\rm vir}$ being the natural scale
of environment dependence.  This is supported by various observational
studies which  have shown that the  environment dependence of galaxies
seems to be restricted to scales $R<R_{\rm vir}$ (e.g., Mo \etal 2004;
Kauffmann \etal 2004; Quintero \etal 2005; Blanton \etal 2006; Blanton
\& Berlind  2007).  Galaxy properties as function  of halo mass can be
studied using either clustering data or using galaxy group catalogues.
The  former  is  less direct as   it  requires modelling  in  order to
translate an observed correlation function into  a description of halo
occupation statistics.  Nevertheless, a great deal of progress in this
area  has been made in  recent years (e.g., Yang  \etal  2003; van den
Bosch \etal 2003, 2007a; Scranton 2003; Magliochetti \& Porciani 2003;
Zehavi \etal  2005; Tinker  \etal 2005; Collister   \& Lahav 2005;  Li
\etal  2006; Wang  \etal 2007a; Cacciato  \etal   2008).  Galaxy group
catalogues allow  a more  direct  study  of  the relation between  the
properties of galaxies and  their dark matter haloes.   In particular,
it assigns  a halo to each  individual galaxy, whereas  the clustering
method  only assigns  haloes   to  galaxies in  a statistical   sense.
Another  advantage of  using galaxy group   catalogues is  that it  is
straightforward to separate  the  galaxy population into centrals  and
satellites  and    to  compute   their  halo-centric  distances    (in
projection).

In van  den Bosch  \etal (2007b, hereafter  Paper~I), we used  a large
galaxy group catalogue  constructed from the SDSS to  study the impact
of  satellite  specific  transformation  processes.  We  compared  the
colours and concentrations of  satellites galaxies to those of central
galaxies of  the same stellar  mass, adopting the hypothesis  that the
latter  are the  progenitors  of the  former.   On average,  satellite
galaxies were  found to be  redder and more concentrated  than central
galaxies of the same  stellar mass, indicating that satellite specific
transformation processes  do indeed operate.   Central-satellite pairs
that are  matched in  both stellar mass  and colour, however,  show no
average concentration  difference, indicating that  the transformation
mechanisms operating on satellites affect colour more than morphology.
We  also  showed that  the  colour  and  concentration differences  of
matched central-satellite pairs are completely independent of the mass
of  the  host halo  (not  to  be confused  with  the  subhalo) of  the
satellite  galaxy, indicating  that satellite  specific transformation
mechanisms are  equally efficient  in host haloes  of all  masses.  In
this  paper we  extend  this  study by  investigating  the ecology  of
satellite galaxies. In particular,  we investigate how the colours and
concentrations of satellite galaxies are correlated with their stellar
mass, their  halo mass, and  their (projected) halo-centric  distance. 
We  believe that  this  is  the optimal  strategy  to investigate  the
effects  of  satellite-specific   transformation  mechanisms  such  as
strangulation, ram-pressure  stripping and harassment.   Throughout we
adopt  a   flat  $\Lambda$CDM  cosmology   with  $\Omega_m=0.238$  and
$\Omega_{\Lambda} =  0.762$ (Spergel \etal 2007) and  we express units
that  depend   on  the   Hubble  constant  in   terms  of   $h  \equiv
H_0/100\kmsmpc$.

\section{Data}
\label{sec:data}

The analysis presented  in this paper is based on  the SDSS DR4 galaxy
group  catalogue of  Yang  \etal (2007;  hereafter  Y07).  This  group
catalogue is constructed applying  the halo-based group finder of Yang
\etal (2005a) to the  New York University Value-Added Galaxy Catalogue
(NYU-VAGC;  see  Blanton \etal  2005),  which  is  based on  SDSS  DR4
(Adelman-McCarthy  \etal 2006).   From  this catalogue  we select  all
galaxies  in  the Main  Galaxy  Sample  with  an extinction  corrected
apparent magnitude  brighter than $r=18$, with redshifts  in the range
$0.01  \leq z \leq  0.20$ and  with a  redshift completeness  $\calC >
0.7$.   As  described  in Y07,  we  use  this  sample of  galaxies  to
construct three group samples: Sample  I, which only uses the $362356$
galaxies with measured  redshifts from the SDSS, Sample  II which also
includes $7091$ galaxies with SDSS photometry but with redshifts taken
from alternative surveys, and  Sample III which includes an additional
$38672$  galaxies that lack  a redshift  due to  fiber-collisions, but
which we  assign the  redshift of its  nearest neighbour  (cf.  Zehavi
\etal  2002).   Unless   specifically  stated  otherwise  the  present
analysis  is based  on Sample  II, which  consists of  369447 galaxies
distributed over 301237 groups.

The  magnitudes and colours of all  galaxies are based on the standard
SDSS petrosian technique (Petrosian  1976;  Strauss \etal 2002),  have
been corrected for  galactic extinction (Schlegel, Finkbeiner \& Davis
1998), and have been  $K$-corrected and evolution corrected (hereafter
$E$-corrected) to $z=0.1$,   using the  method described  in  (Blanton
\etal 2003a).  We    use the  notation  $^{0.1}M_r$ to   indicate  the
resulting absolute  magnitude in the $r$-band.   In addition, for each
galaxy we  compute a stellar mass,  $M_*$, using the relations between
stellar  mass-to-light ratio and colour of  Bell  \etal (2003; see Y07
for details).  In addition to the magnitude and  stellar mass, we also
use the    $^{0.1}(g-r)$  colour  of  each   galaxy  as   well  as its
concentration $C=r_{90}/r_{50}$.   Here $r_{90}$  and $r_{50}$ are the
radii that contain 90  and 50 percent of  the Petrosian $r$-band flux,
respectively.  In what follows, we  interpret the $^{0.1}(g-r)$ colour
as  an indicator of the  galaxy's star formation history (mean stellar
age),  though  we acknowledge that because   of  dust attenuation this
connection  may not  be  very tight.  In   addition, we  interpret the
concentration parameter   as   reflecting  the galaxy's  mass-assembly
history.  As shown by Strateva \etal (2001), $C$ is a reasonable proxy
for Hubble   type,   with  $C>2.6$ corresponding  to    an  early-type
morphology. The  link to the mass-assembly history  owes to the common
belief that a galaxy's   morphology (in particular  its  disk-to-bulge
ratio), is    related to the importance  of   major mergers (and other
violent disturbances) during its mass assembly.

For  each group in  our  catalogue we have two   estimates of its dark
matter   halo mass $M_h$:    one based on   the  ranking of its  total
characteristic  luminosity, and the other based  on the ranking of its
total characteristic stellar mass.  As shown in  Y07, both halo masses
agree  very   well with  each  other,  with   an average  scatter that
decreases from $\sim  0.1$ dex at the low  mass end to $\sim 0.05$ dex
at  the  massive end.  In  addition,  detailed tests  with mock galaxy
redshift catalogues have  demonstrated that our  group masses are more
reliable than  those based  on  the velocity dispersion  of the  group
members (Yang  \etal 2005b;    Weinmann \etal 2006a;   Y07).   Further
support for the accuracy of our group masses comes from an analysis of
their clustering properties   (Wang \etal 2007b).    In this paper  we
adopt   the group masses based  on  the stellar  mass ranking. We have
verified, though, that none of our  results change significantly if we
adopt    the    luminosity-rank     based    masses    instead    (see
\S\ref{sec:groupmass})

Finally,  for each  group  member we  determine  the projected  radius
$R_{\rm  proj}$ from the  luminosity weighted  group center  using the
angular separation  and the angular diameter distance  at the redshift
of   the  group.    We  normalize   these  projected   radii   by  the
characteristic radius of the group, $R_{180}$, which is defined as the
radius inside of which the  dark matter halo associated with the group
has an average overdensity of $180$.  This radius is computed from the
mass and redshift of the group using equation~(5) in Y07.

\subsection{Sample definition}
\label{sec:satsample}

From Sample II we select  all satellite galaxies, defined as all group
members that are not the most  massive group member, of groups with an
assigned  halo mass  in the  range $12.0  \leq \log[M_h/(h^{-1}\Msun)]
\leq 15.0$.  In addition, the  satellites are required to have stellar
masses in  the range $9.0 \leq \log[M_*/(h^{-2}\Msun)]  \leq 11.5$ and
normalized projected radii $R_{\rm proj}/R_{180} \leq 1.0$. Neither of
these criteria is particularly  restrictive, so that the vast majority
($\sim 88\%$) of all satellite galaxies in our group catalogue make it
into the  sample used  here.  The resulting  sample consists  of 60245
satellite galaxies,  which we use to study  the interrelations between
colour,  concentration,   halo  mass,  stellar   mass,  and  projected
halo-centric radius.

Note  that this  is not  a volume-limited  sample.   Consequently, the
sample suffers from Malmquist  bias, causing an artificial increase of
the  average  luminosity (and  also  stellar  mass)  of the  satellite
galaxies  with increasing  redshift.   To correct  for  this bias,  we
weight  each galaxy  by $1/V_{\rm  max}$, where  $V_{\rm max}$  is the
comoving volume  of the Universe out  to a comoving  distance at which
the galaxy would still have made the selection criteria of our sample.
In  what follows all  distributions are  weighted by  $1/V_{\rm max}$,
unless specifically stated otherwise
\begin{figure}
\centerline{\psfig{figure=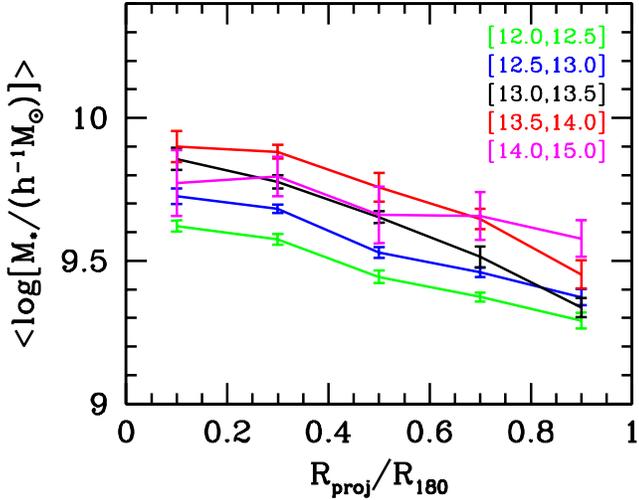,width=\hssize}}
\caption{The average satellite mass as a function of the normalized,
  projected radius from the  group center.  Different lines correspond
  to  different  bins in  halo  mass,  as  indicated. Two  trends  are
  evident.   First of  all,  at a  given  $R_{\rm proj}/R_{180}$  more
  massive  haloes  contain,  on  average,  more  massive  satellites.  
  Secondly, there  is clear evidence  for mass segregation,  with more
  massive  satellites (in  a  given halo  mass),  residing at  smaller
  (projected) radii.}
\label{fig:mass_seg}
\end{figure}

\section{Results}
\label{sec:res}

The  main  aim of   this  paper  is  to  investigate  the  conditional
probability functions  $P(^{0.1}(g-r)  \vert M_*, M_h,  R_{\rm proj})$
and $P(C \,  \vert M_*, M_h, R_{\rm proj})$  of satellite galaxies. In
particular,  we  wish to  establish  which of the conditionals, $M_*$,
$M_h$ or $\rp$,  is most relevant  for setting the colours (related to
the  star  formation histories)  and  concentrations  (related  to the
mass-assembly histories) of satellite galaxies.

In order  to   facilitate the  interpretation of   these   probability
functions and their  moments  we  first  focus  on the  interrelations
between  the   three  conditionals.  These  are  nicely  summarized in
Fig.~\ref{fig:mass_seg}, which plots  the average stellar mass  of the
satellite  galaxies,   $\langle  \log[M_*/(h^{-2}\Msun)] \rangle$,  as
function of $\rp$  for different  bins  in halo mass.   Two trends are
clearly apparent: at fixed $\rp$, the average satellite mass increases
with increasing halo mass, while at fixed $M_h$, the average satellite
mass decreases with increasing $\rp$.  The former simply reflects that
the   characteristic  mass  (or  luminosity)  of    satellite galaxies
increases  with  increasing halo mass  (cf.,  Yang  \etal 2005c; Zheng
\etal 2005; Skibba,  Sheth \& Martino 2007;  Yang, Mo \& van den Bosch
2008).   The latter    indicates that  the   spatial distribution   of
satellite galaxies  in dark   matter  haloes has undergone  some  mass
segregation, and  is    consistent with  the   luminosity  segregation
observed  in galaxy  clusters (e.g., Rood  \&  Turnrose 1968; Quintana
1979;  den  Hartog \& Katgert  1996;   Adami, Biviano \&  Mazure 1998;
Lares, Lambas \& S\'anchez 2004; McIntosh \etal 2005).  As we will see
below, these two  trends are essential  for understanding the  various
relations between colour  and concentration on  the one hand and  halo
mass, stellar mass, and halo-centric radius on the other.

\subsection{First Moments}
\label{sec:averages}

\begin{figure*}
\centerline{\psfig{figure=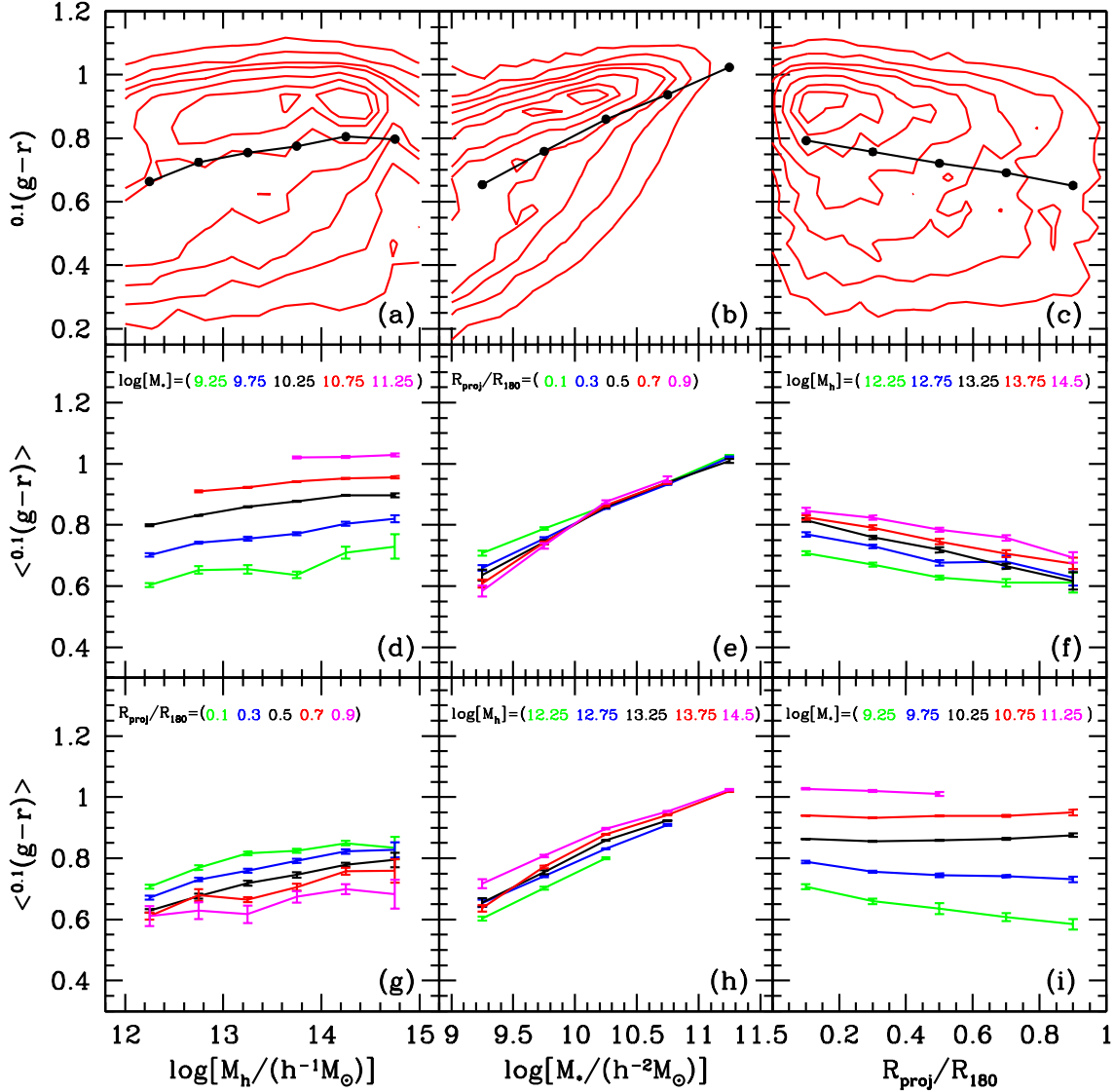,width=0.85\hdsize}}
\caption{The  contours  in  the  upper  panels  reflect the  $1/V_{\rm
    max}$-weighted distributions  of satellite colours as functions of
  halo mass $M_h$ (left), stellar mass  $M_*$ (middle) and normalized,
  projected   radius   $\rproj$ (right).   The   connected  solid dots
  indicate   the average  satellite   colour   as function   of  these
  quantities.  The  lower six panels show the  average colour again as
  function of these  three quantities, but this time  split in bins of
  one of  the other two quantities,  as indicated.   The values at the
  top   of each panel indicate  the  central values  of  each bin.  In
  particular,  for  $\rproj$ we  consider  bins  of $[0.0,0.2]$ (green
  lines),  $[0.2,0.4]$   (blue  lines),  $[0.4,0.6]$   (black  lines),
  $[0.6,0.8]$ (red lines)  and  $[0.8,1.0]$ (magenta lines).   For the
  stellar mass  we    adopt  bins  in  $\log[M_*/(h^{-2}\Msun)]$    of
  $[9.0,9.5]$  (green lines), $[9.5,10.0]$ (blue lines), $[10.0,10.5]$
  (black lines), $[10.5,11.0]$  (red lines) and $[11.0,11.5]$ (magenta
  lines).   And   finally,   for   the halo  mass  we   use    bins in
  $\log[M_h/(h^{-1}\Msun)]$ of    $[12.0,12.5]$     (green     lines),
  $[12.5,13.0]$   (blue      lines), $[13.0,13.5]$     (black  lines),
  $[13.5,14.0]$ (red lines)  and $[14.0,15.0]$ (magenta lines).   Note
  that the last  bin in $M_h$  is twice  as broad  as the other  bins,
  which is done to assure a sufficient number of satellite galaxies to
  be able to   calculate  reliable statistics.   Errorbars  (sometimes
  barely   visible) indicate the  standard  variance  obtained from 20
  jackknife   samples.  Note how   the   average colours  of satellite
  galaxies at fixed  stellar mass are  virtually independent  of their
  environment (halo mass and halo-centric radius).}
\label{fig:aver_col}
\end{figure*}

As a first step in our investigation of $P(^{0.1}(g-r) \vert M_*, M_h,
R_{\rm proj})$ and $P(C \, \vert M_*, M_h, R_{\rm proj})$ of satellite
galaxies  we focus  on their  first  moments.  Fig.~\ref{fig:aver_col}
shows the  relations between the  $\col$ colour of  satellite galaxies
and  their  halo mass  $M_h$,  their  stellar  mass $M_*$,  and  their
normalized, projected,  halo-centric radius, $\rp$.   The upper panels
show contour plots of the various distributions, while the solid lines
reflect  the   average  colour.  Panel  (a)  shows   that  the  colour
distribution of satellite galaxies has a remarkably weak dependence on
halo  mass: over the  entire range  of halo  masses probed  the colour
distribution  is clearly skewed,  with a  relatively narrow  peak near
$^{0.1}(g-r)=0.85$ and an extended wing  to bluer colours.  There is a
weak trend, though, that the  average satellites are somewhat bluer in
less massive haloes.

The colour-stellar  mass distribution, shown  in panel (b),  reveals a
pronounced and narrow  red sequence which is clearly  tilted with more
massive  red   sequence  satellites  being  redder.    In  fact,  this
colour-stellar mass relation of  satellite galaxies looks very similar
to that for the entire  galaxy population (cf. Fig.~7 in Paper~1).  We
emphasize that this  is not a trivial result  since satellite galaxies
only  contribute  between 20  and  40  percent  of the  entire  galaxy
population  (see  \S\ref{sec:intro}).   As  discussed in  Paper~I,  it
implies that  centrals and  satellites of the  same stellar  mass have
very   similar  colour   distributions,  which   already   puts  tight
constraints  on  the efficiencies  with  which  the various  satellite
specific transformation processes operate.

Panel  (c)  of   Fig.~\ref{fig:aver_col}  shows  the  distribution  of
satellite galaxies as function of  $\col$ and $\rproj$.  First of all,
the number  of satellites seems  to peak around $\rp=0.15$;  the rapid
decline  toward smaller  radii, however,  is an  artefact of  the data
sample,  and owes  to the  problem of  fiber collisions  in the  SDSS. 
Since galaxies  lost from  the survey due  to fiber collisions  do not
have any specific colours or concentrations, this has no impact on our
results, as we demonstrate in \S\ref{sec:fibcoll}.  The steady decline
in the  number of  satellites at larger  radii is genuine,  and simply
reflects the fact that the number density of satellites decreases with
increasing halo-centric  radius faster  than $r^{-1}$ (e.g.,  Beers \&
Tonry  1986; Carlberg,  Yee \&  Ellingson 1997a;  van der  Marel \etal
2000;  Lin, Mohr  \& Stanford  2004; van  den Bosch  \etal  2005; Chen
2007).  Finally, there  is also a clear indication  that satellites at
larger  halo-centric radii  are, on  average, bluer.   Again,  this is
consistent  with numerous  other  studies (e.g.,  Biviano \etal  1996;
Colless \& Dunn 1996; Carlberg \etal 1997b; Lares \etal 2004).
\begin{figure*}
\centerline{\psfig{figure=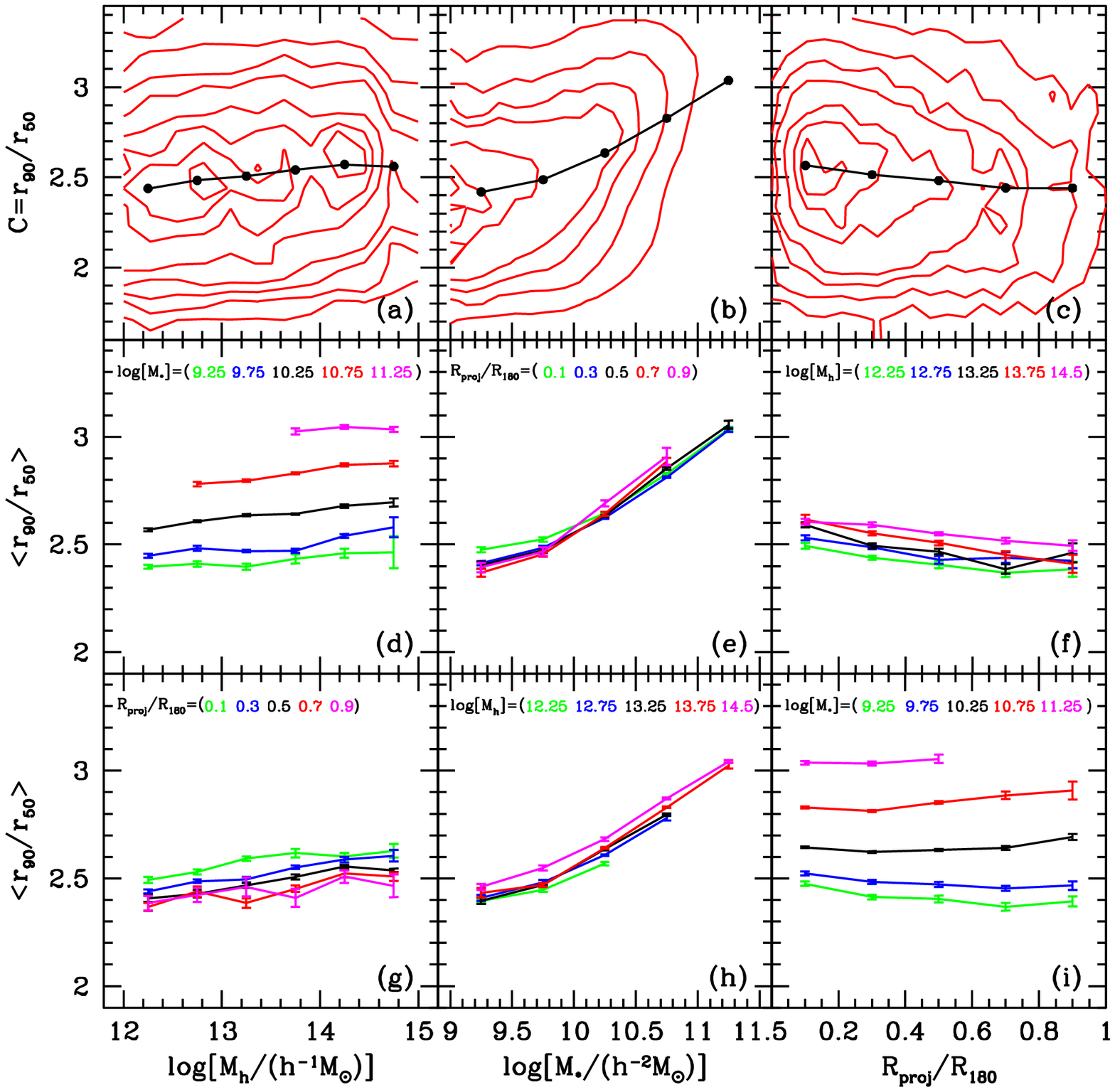,width=0.85\hdsize}}
\caption{Same as  Fig.~\ref{fig:aver_col}  but for  the  concentration
  parameter $C=r_{90}/r_{50}$  rather than  the colour. Note  that the
  overall  trends are  remarkably  similar.  In  particular, at  fixed
  stellar  mass the  average  concentration of  satellite galaxies  is
  virtually   independent  of   their  environment   (halo   mass  and
  halo-centric radius).  See text for a detailed discussion.}
\label{fig:aver_conc}
\end{figure*}

The  lower  six panels  of  Fig.~\ref{fig:aver_col}  show the  average
$\col$  colour  of  satellite  galaxies  as function  of  $M_h$  (left
column), $M_*$ (middle column)  and $\rproj$ (right column).  Lines of
different colours correspond to different bins in one of the other two
parameters, as indicated  at the top of each  panel.  Errorbars (often
barely  visible) reflect  the standard  deviations obtained  using the
jackknife technique. To  that extent we divide the  group catalogue in
$N=20$  subsamples of  roughly equal  size, and  recalculate  the mean
colour 20 times, each time leaving  out one of the 20 subsamples.  The
jackknife estimate of the standard deviation then follows from
\begin{equation}
\label{jack}
\sigma_x = \sqrt{{N-1 \over N} \sum_{i=1}^N \left(x_i - \bar{x}\right)^2}
\end{equation}
with $x_i$ the average colour obtained from jackknife sample $i$.

Panel (d) of Fig.~\ref{fig:aver_col}  shows that at fixed stellar mass
the average colour  of satellite galaxies depends only  very weakly on
halo mass.   On the other  hand, at fixed  halo mass, there is  a very
strong dependence on stellar  mass, with more massive satellites being
redder.  The  same trend is also  visible from panel  (h), which shows
the average colour as a function of stellar mass for different bins in
halo mass.  Panels (f) and  (g) show that the average satellite colour
becomes slightly  bluer with  increasing halo-centric radius  at fixed
halo mass.   At the same  time, satellites at  a fixed $\rp$  are also
somewhat redder if they reside in more massive haloes.  However, as is
evident from panel (e), at  fixed stellar mass there is no discernable
dependence of the average satellite colour on $\rp$.
\begin{figure*}
\centerline{\psfig{figure=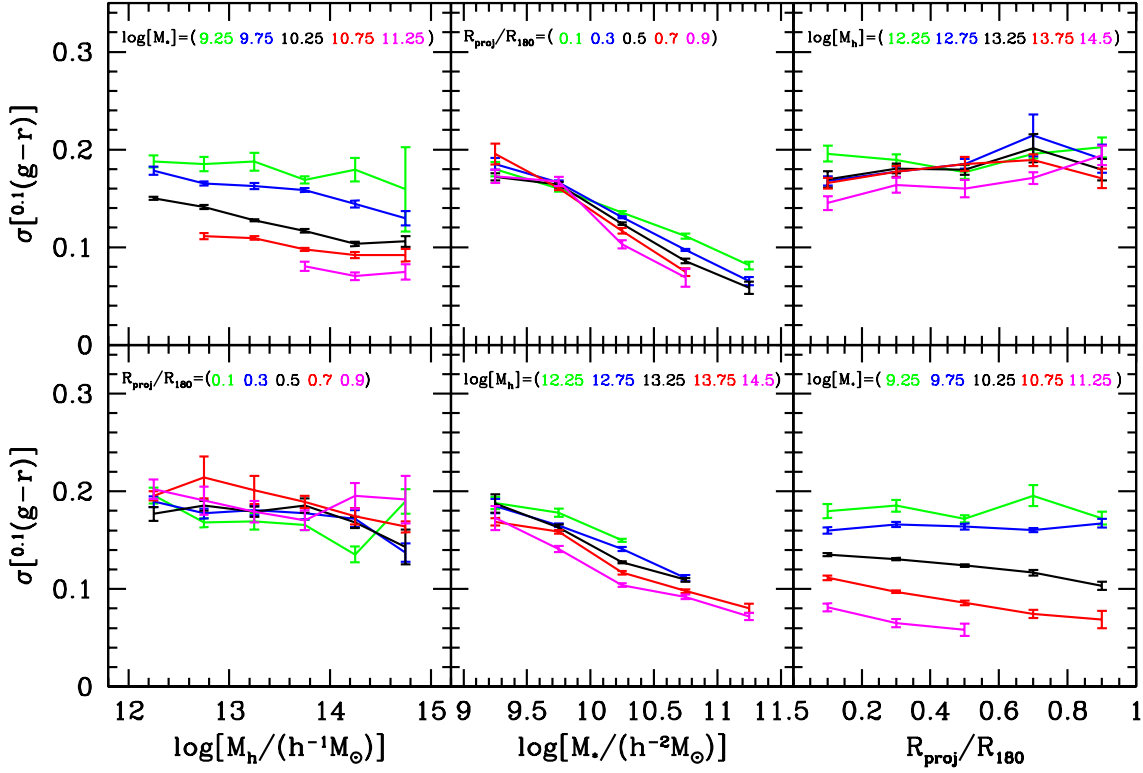,width=0.85\hdsize}}
\caption{The  standard  deviations $\sigma[^{0.1}(g-r)]$ of  the colour
  distribution of  satellite galaxies as  function of halo  mass (left
  panels), stellar  mass (middle  panels) and projected  radius (right
  panels).  Lines of different  colour correspond to different bins in
  one of  the other three  conditionals, as indicated.  The  values at
  the top of  each panel indicate the central values  of each bin, and
  the  colour  coding  is  the  same as  in  Fig.~\ref{fig:aver_col}.  
  Errorbars reflect the standard deviations obtained from 20 jackknife
  samples.}
\label{fig:sigm_col}
\end{figure*}

Fig.~\ref{fig:aver_conc} shows the same as Fig.~\ref{fig:aver_col} but
now for  the average concentrations of the  satellite galaxies, rather
than  their colours.  Note  that the  overall behavior  is remarkably
similar to  the case of  the average colours:  (i) at fixed  halo mass
there is a strong dependence  on stellar mass (more massive satellites
are more centrally concentrated), while at fixed stellar mass there is
almost no  halo mass dependence;  (ii) at fixed  halo mass there  is a
weak   (barely   significant)   dependence  on   halo-centric   radius
(satellites at larger $\rp$ are less concentrated), and at fixed $\rp$
there is  a similarly weak dependence  on $M_h$; (iii)  at fixed $\rp$
satellite  concentrations depend  strongly on  stellar mass,  while at
fixed stellar mass there is no discernable dependence on $\rp$.
 
To  summarize, the  average  colours and  concentrations of  satellite
galaxies  show almost  no sign  of intrinsic  environment  dependence. 
Rather, they seem to be completely determined by their stellar masses:
to good  accuracy Figs.~\ref{fig:aver_col} and~\ref{fig:aver_conc} can
be  used  to  read  off  the average  colours  and  concentrations  of
satellite galaxies of a given stellar mass {\it without having to know
  anything about  their environment}.  At least a  significant part of
the (already weak)  dependence on halo mass is  not causal, but merely
owes  to  the fact  that  more  massive  haloes contain  more  massive
satellites   (cf.   Fig.~\ref{fig:mass_seg}).   Similarly,   the  mild
dependence  on   $R_{\rm  proj}/R_{180}$   seems  to  owe   mainly  to
mass-segregation, rather than to  a causal relation between colour and
halo-centric radius.  This remarkable dearth of environment dependence
for  satellite galaxies  is the  main results  of this  paper.   As we
discuss  in  \S\ref{sec:concl},   it  has  profound  implications  for
theories  of  galaxy  formation.   The  remainder  of  this  paper  is
therefore  concerned with  addressing the  robustness  and statistical
significance  of  these  findings,  and  with  reconciling  them  with
previous, at first sight antithetical, results.

\subsection{Higher order moments}
\label{sec:higher}

The above  analysis, based on the average  colours and concentrations,
suggests  that  the  star  formation  histories  and  morphologies  of
satellite  galaxies are  virtually independent  of their  environment. 
Instead, they  seem to  be governed almost  entirely by  their stellar
masses.  However, the  averages only reflect the first  moments of the
full probability  distributions $P(^{0.1}(g-r) \vert  M_*, M_h, R_{\rm
  proj})$  and $P(C  \,  \vert  M_*, M_h,  R_{\rm  proj})$. Since  two
distributions can be very different and yet have the same average, the
above conclusion is only valid for the first moments of the colour and
concentration distributions of satellite galaxies.

\begin{figure*}
\centerline{\psfig{figure=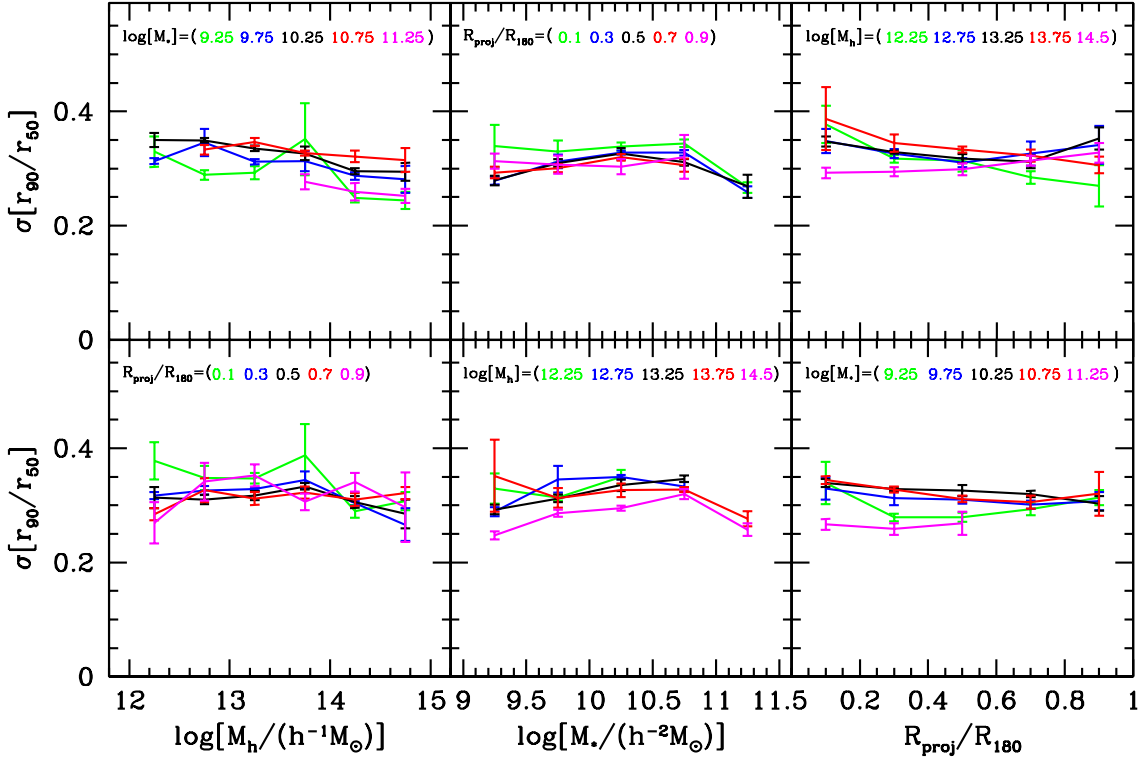,width=0.85\hdsize}}
\caption{Same as Fig.~\ref{fig:sigm_col} but for the concentrations
  of satellites rather than their colour.}
\label{fig:sigm_conc}
\end{figure*}

As a logical next step, we therefore  now focus on the second moments.
Fig.~\ref{fig:sigm_col}      shows     the     standard     deviations
$\sigma[^{0.1}(g-r)]$  as functions of  the  three  conditionals.  The
overall behavior is  very similar as  for  the averages: (i)  at fixed
halo mass there is a strong dependence on stellar mass, while at fixed
stellar mass, the dependence  on  halo mass  is much weaker,  (ii)  at
fixed  $\rp$ there is  a  strong  dependence on  $M_*$,  but at  fixed
stellar  mass  there is virtually no  dependence  on projected radius.
Thus the  second moments   of  $P(^{0.1}(g-r) \vert  M_*, M_h,  R_{\rm
  proj})$ depend  most strongly on  $M_*$, with only a weak dependence
on    $M_h$,    and     virtually      no   dependence    on    $\rp$.
Fig.~\ref{fig:sigm_conc} shows the    same but now  for  the satellite
concentrations.  Contrary to  the colours, the standard  deviations of
$P(C \vert M_*, M_h, R_{\rm  proj})$ seem to  be almost independent of
stellar  mass.  In   fact,  within the   errors,  there  is  no really
significant dependence on any of the three conditionals.

We can therefore  conclude that the statement that  the star formation
histories  and  morphologies  of  satellite galaxies  are  (virtually)
independent  of  their environment  remains  valid  up  to the  second
moments of their respective distribution functions.

\begin{figure*}
\centerline{\psfig{figure=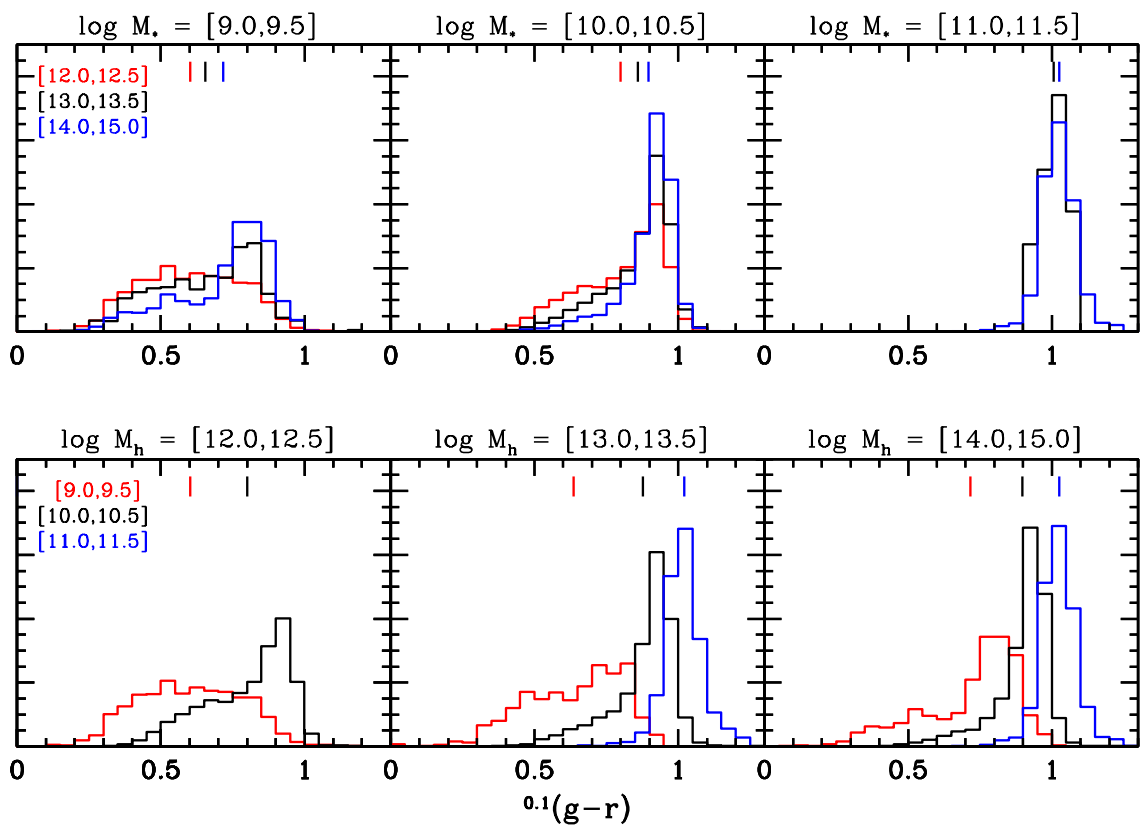,width=0.85\hdsize}}
\caption{Histograms of    the colour distributions  of   satellites for
  various bins  in halo  mass and stellar  mass.  In the  upper panels
  different colours correspond to  satellites of the same stellar mass
  (indicated  at the  top of  each panel)  but residing  in  groups of
  different halo mass (indicated by the values in square brackets). In
  the  lower  panels different  colours  correspond  to satellites  in
  haloes of the same mass (indicated  at the top of each panel) but of
  different stellar mass (indicated by the values in square brackets).
  The small  vertical bars near the  top of each  panel, indicated the
  averages of the corresponding distributions. See text for a detailed
  discussion.}
\label{fig:his_col}
\end{figure*}

Expanding this investigation, we  could next present similar plots for
the third moments, the fourth  moments, etc.  However, already for the
third moment (or equivalently, the  skewness) we find that the results
become too  noisy (i.e., the jackknife  errors are too  large) to make
any quantitative  assessment.  Therefore, we choose to  show (a subset
of) the  full probability distributions instead.  The  upper panels of
Fig.~\ref{fig:his_col} show the full colour distributions of satellite
galaxies of a given stellar mass  (indicated at the top of each panel)
for  three different bins  in halo  mass (indicated  by the  values in
square  brackets).  The  lower panels,  on  the other  hand, show  the
colour distributions of satellites in haloes of a given mass for three
different bins in stellar mass.   The small vertical bars near the top
of   each   panel  indicate   the   averages   of  the   corresponding
distributions.   Only  two  distributions   are  shown  in  the  upper
right-hand  and lower  left-hand panels,  since haloes  with  $12 \leq
\log[M_h/(h^{-1}\Msun)] \leq  12.5$ do not  contain satellite galaxies
with $\log[M_*/(h^{-2}\Msun)] \geq 11$.  As is clear from a comparison
of the upper  and lower panels, the colour  distributions of satellite
galaxies are  far more dependent  on stellar mass  than on halo  mass. 
However, it is also clear  that satellite galaxies of the same stellar
mass  that   reside  in  haloes   of  different  masses   have  colour
distributions  that are  not exactly  the  same.  In  fact, there  are
significant differences, though they  typically relate to higher order
moments of the  distributions.  For example, in the  upper left panel,
the distributions for  $\log[M_h/(h^{-1}\Msun)] \geq 13.0$ are clearly
negatively  skewed, while  that for  $12  \leq \log[M_h/(h^{-1}\Msun)]
\leq 12.5$  has a positive  skewness.  We therefore conclude  that the
dependence of  satellite colour on environment is  extremely mild, and
only   apparent  from   the   higher-order  moments   of  the   colour
distributions.

\begin{figure*}
\centerline{\psfig{figure=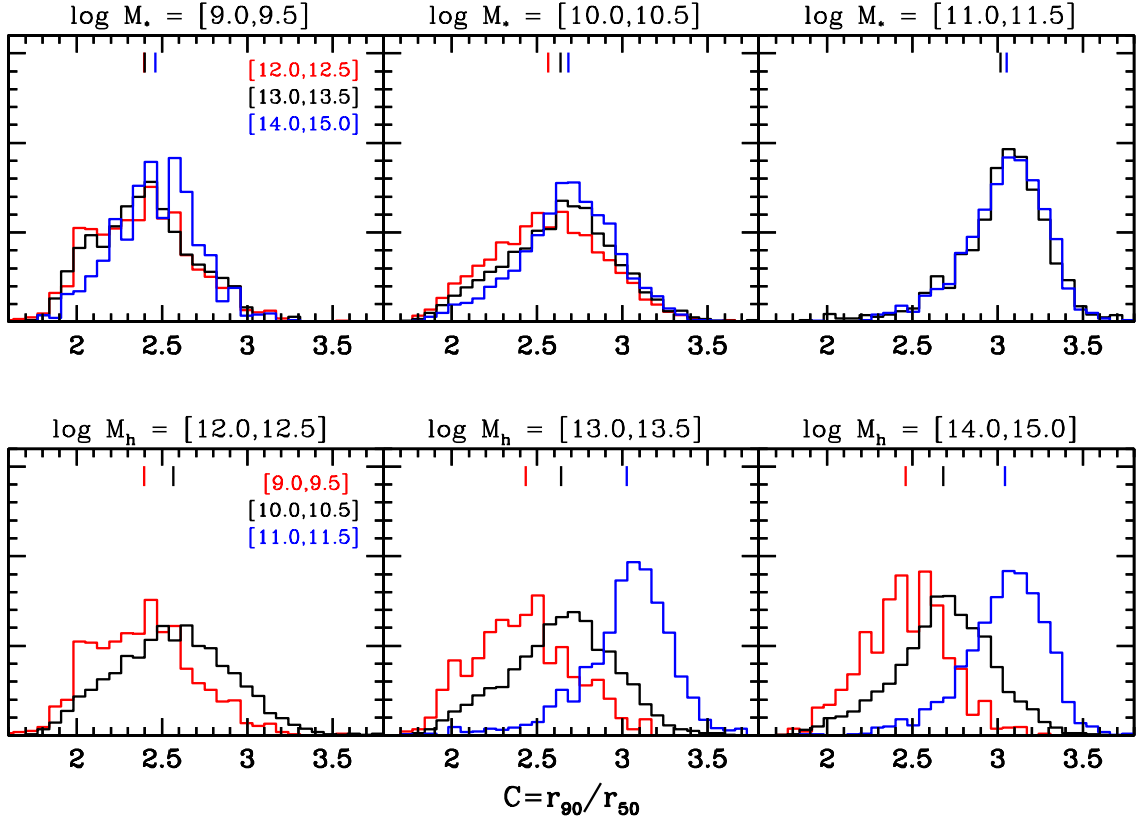,width=0.85\hdsize}}
\caption{Same as Fig.~\ref{fig:his_col} but for the concentrations
  of satellites rather than their colour.}
\label{fig:his_conc}
\end{figure*}

Finally, Fig.~\ref{fig:his_conc} shows  the distributions of satellite
concentrations for the  same bins in halo mass and  stellar mass as in
Fig.~\ref{fig:his_col}.   The  conclusions are  the  same  as for  the
colours:  at fixed  stellar mass  the distributions  only  depend very
weakly on halo  mass (even more weakly than in the  case of the colour
distributions),  while at  fixed halo  mass, the  distributions depend
very strongly on stellar mass.

\section{Testing the Robustness}
\label{sec:tests}

In order to check  the robustness of the  above results, we perform  a
number of tests, which we now describe in turn.

\subsection{Group masses}
\label{sec:groupmass}

As mentioned in  \S\ref{sec:data} each group in  our catalogue has two
estimates for  the mass  of its corresponding  dark matter  halo:  one
based on its ranking of the  characteristic luminosity, defined as the
total luminosity of  all group members with $^{0.1}M_r  - 5\log h \leq
-19.5$,  and one based on   its ranking of  the characteristic stellar
mass, defined as  the  total stellar  mass  of all group members  with
$^{0.1}M_r - 5\log h \leq -19.5$  (see Y07 for  details).  Thus far we
have used the  latter.  Since the  stellar masses are determined  from
the colours of the  galaxies,  one could  potentially be worried  that
this has somehow influenced the dependence of satellite colour on halo
mass.  To test   this, we repeat our   analysis using the  halo masses
based  on the luminosity ranking.  We  also change the definition of a
satellite  galaxy:  previously a satellite  galaxy  was defined as any
group member  that  is not the most   massive,  while now  a satellite
galaxy is any group  member that is  not the most  luminous. In  a few
cases this  turns a satellite galaxy  into a central  galaxy, and vice
versa.  The left-hand column of Fig.~\ref{fig:tests} shows the results
thus obtained,  and which should be  compared to the  middle column of
Fig.~\ref{fig:aver_col}.  Since we   only consider satellite  galaxies
that reside  in groups with an  assigned halo mass  in the range $12.0
\leq \log[M_h/(h^{-1}\Msun)] \leq 15.0$, the satellite samples in both
cases are not exactly  the same. In  fact, when using the group masses
based  on  the  luminosity   ranking we  obtain   a  sample of   60835
satellites, compared to 60245 in the case of the stellar mass ranking.
This  difference, however, is extremely  small, so  that the upper two
panels in  the  first    column of Fig.~\ref{fig:tests}    are  almost
identical  to those in  the  middle column of Fig.~\ref{fig:aver_col}.
The more interesting comparison  regards  the  lower panels in   these
columns, where  the average colours  as function  of stellar mass  are
shown  for different bins  in halo mass.   As can be seen, the results
are almost  indistinguishable,   indicating   that our    results  are
extremely robust  to whether   we  use the halo   masses based  on the
luminosity ranking or the stellar mass  ranking. We have verified that
this is  also the  case for all  other   statistics presented  in this
paper.

\subsection{Fiber Collisions}
\label{sec:fibcoll}

\begin{figure*}
\centerline{\psfig{figure=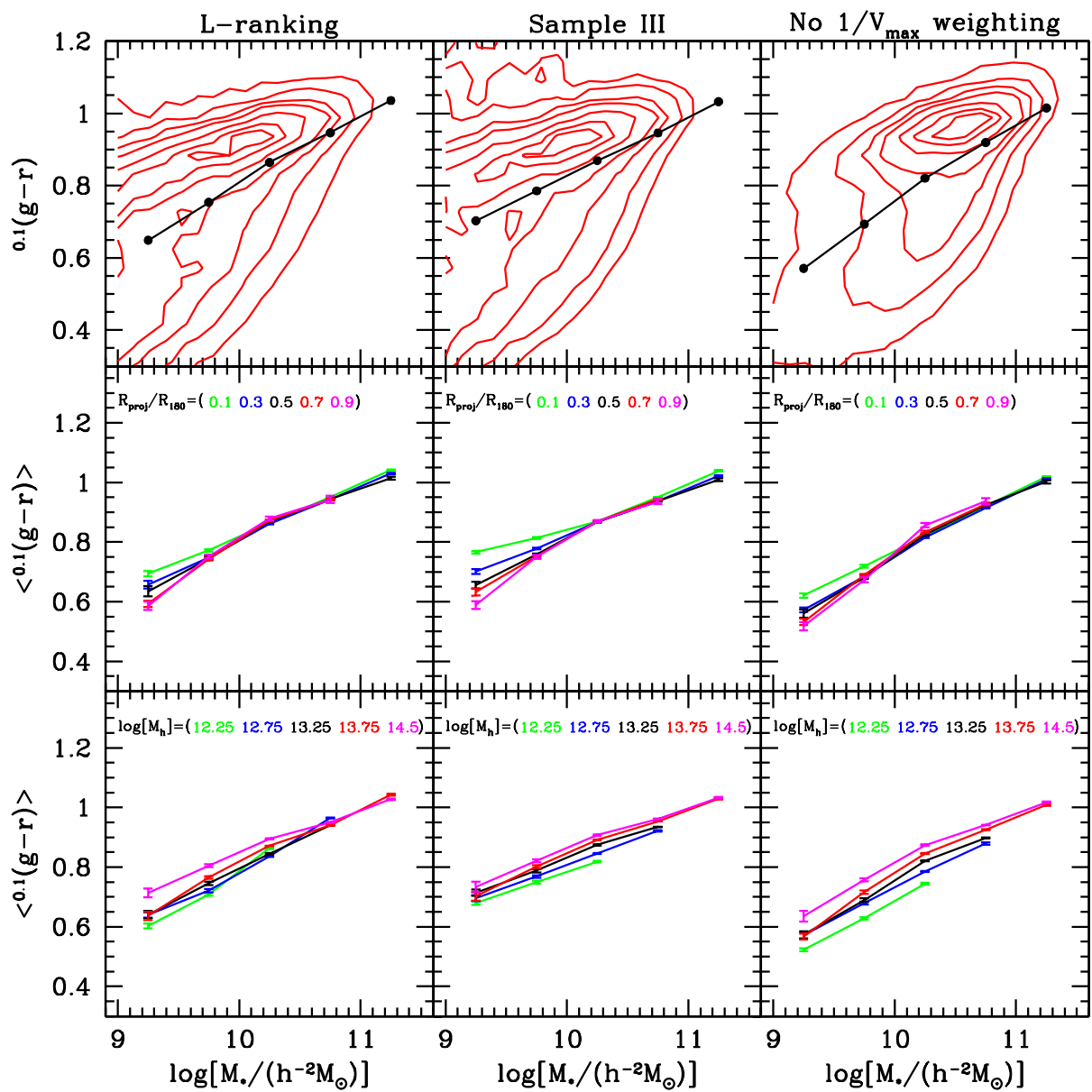,width=0.9\hdsize}}
\caption{Same as the middle column of Fig.~\ref{fig:aver_col} but using
  halo masses based on the luminosity ranking, rather than the stellar
  mass ranking (left-hand column), using Sample III rather than Sample
  II  (middle   column),  and  using  no   $1/V_{\rm  max}$  weighting
  (right-hand column).  In all  cases, the average colours as function
  of stellar  mass are very  similar, indicating that our  results are
  robust to  uncertainties in halo  mass, incompleteness due  to fiber
  collisions, and  inaccuracies in the correction for  Malmquist bias. 
  In particular, in all cases the average colour of satellite galaxies
  depends strongly on  stellar mass with no or  very little dependence
  on environment (halo-centric  radius, as shown in the  middle row of
  panels, or  halo mass, shown  in the lower  panels). See text  for a
  detailed discussion.}
\label{fig:tests}
\end{figure*}

As discussed in Y07,  the galaxy sample  on  which group Sample II  is
based  suffers  from incompleteness due to    fiber collisions: no two
fibers on the same SDSS plate can be closer than  55 arcsec.  Although
this  fiber collision constraint is partially   alleviated by the fact
that neighboring plates have overlap regions, $\sim  7$ percent of all
galaxies eligible   for spectroscopy do not  have  a measured redshift
(Blanton \etal 2003d).  Since  fiber  collisions are more  frequent in
regions of high (projected) density, they  are more likely to occur in
richer  groups, thus  causing a systematic  bias.    In Sample III  we
(partially) correct  for  this  by  assigning galaxies that   lack  an
observed redshift due to  fiber collisions (hereafter  fiber-collision
galaxies) the redshift of the galaxy with which it collided.  As shown
in Zehavi  \etal (2002), roughly  60  percent of  the  fiber-collision
galaxies have a redshift within $500  \kms$ of their nearest neighbor,
which justifies the above procedure.  However, there are also cases in
which the   fiber-collision galaxy has a  true  redshift  that is very
different from that of  its nearest neighbor.  In  this case the above
procedure will  cause the  group    finder to falsely  identify    the
fiber-collision galaxy as a group member (and  with the wrong absolute
magnitude and  stellar mass).  In  order to avoid these  failures from
impacting  on our  results,  we have  used Sample  II as our  fiducial
sample  in the present   study (see  \S\ref{sec:data}).  Nevertheless,
since the above method seems to work well for $\sim 60$ percent of the
fiber-collision galaxies, it is worthwhile to check if our results are
sensitive to whether  we include fiber-collision  galaxies or not.  We
therefore  repeat    our analysis using  Sample  III.    The selection
criteria  discussed in  \S\ref{sec:satsample} now   yield a sample  of
92546 satellite galaxies  (32301 more  than for  Sample II). Yet,  the
results, shown  in the middle    column of Fig.~\ref{fig:tests},   are
almost  identical  to those obtained  from Sample   II: once again, at
fixed stellar mass the average satellite  colour is almost independent
of  $M_h$ and $\rp$.  The  main difference  with  respect to Sample II
regards the tails of  the colour-stellar mass  distribution, as can be
seen from the contour plot in the upper panel.  In Sample III there is
a  tail of  faint, very  red galaxies, which   is absent in Sample II.
These galaxies are most likely  fiber-collision galaxies for which the
assigned redshift has a large error.  Another difference, which is not
shown here, is that in Sample III there is no decline in the number of
satellites for  $\rp  \lta  0.15$  as in  the  upper right  panels  of
Figs.~\ref{fig:aver_col} and~\ref{fig:aver_conc}.  However,  since the
fiber-collision galaxies do not have any preferred colour, this has no
impact on the average  colour  of satellite  galaxies at these   small
radii.

\subsection{Malmquist Bias}
\label{sec:malmquist}

As mentioned in     \S\ref{sec:satsample}, the sample  of  groups  and
galaxies used  for our analysis  is  flux-limited, not volume-limited.
In  order  to correct  for  the  consequential  Malmquist bias we have
weighted each  galaxy by $1/V_{\rm max}$,  the inverse of the comoving
volume out to which  the galaxy would   have made it into our  sample.
However, in computing  $V_{\rm max}$ we did  not account  for the fact
that the flux-limit of the SDSS varies mildly with the position on the
sky (see  for example Blanton \etal 2003b).    In order to investigate
how sensitive our results are to  the $1/V_{\rm max}$ weighting scheme
adopted,   we  now repeat our  analysis    without any weighting.  The
results are shown  in  the right-hand column of  Fig.~\ref{fig:tests}.
As expected the distribution of  galaxies in the  plane of colour {\it
  vs.}  stellar mass  is significantly different, showing a relatively
larger contribution from massive (and hence  red) galaxies.  As can be
seen from the black solid line in the upper panel, at the low mass end
satellites  are about 0.1 magnitude bluer  on average than in the case
with weighting.  A galaxy of  a given stellar  mass that is bluer will
also typically be brighter, and can therefore be  seen out to a higher
redshift.  In other  words,   by not   applying the  $1/V_{\rm   max}$
weighting one artificially overestimates the fraction of blue galaxies
at a given stellar  mass, thus causing the  average  colour to  be too
blue (but  only mildly  so).  Yet, despite  the fact  that the average
colours at a given  stellar mass are somewhat  different, even in this
extreme  case without any weighting  to correct for Malmquist bias one
finds that there is only a very mild halo  mass dependence of $\langle
^{0.1}(g-r)\rangle$ at fixed  $M_*$  and no significant  dependence on
$\rp$ at   all, in  good  agreement   with  our fiducial   results  in
Fig.~\ref{fig:aver_col}.

For brevity  we only  show the above  tests for the  average satellite
colours as function  of stellar mass. However, the  average colours as
function of halo  mass or halo-centric radius, as  well as the average
concentrations,  all  point  to   the  same  conclusion:  the  results
presented in \S\ref{sec:res} are extremely robust.  They do not depend
on our choice of the group  mass indicator, they are not influenced by
sample  incompleteness  due  to  fiber-collisions, and  they  are  not
affected by an imperfect correction for Malmquist bias.

\section{Comparison with previous studies}
\label{sec:comp}

The  analysis presented  above shows  that at  fixed stellar  mass the
colours  and  concentrations of  satellite  galaxies  are only  weakly
dependent  on  their  environment.   This differs  substantially  from
several previous  studies, which have argued for  a strong environment
dependence of galaxy colours  or star formation indicators (e.g., Hogg
\etal 2004; Kauffmann \etal  2004; Blanton \etal 2005b; Weinmann \etal
2006a)

Our study differs from most of these on the following grounds:
\begin{enumerate}

\item we  have used halo  mass and halo-centric radius  as environment
  indicators, as opposed to a projected number density of galaxies.

\item we have  split our galaxy population in  centrals and satellites
  and only focused on the latter.
  
\item we  have used broad-band colours, rather  than more sophisticated
  star formation activity indicators,  such as the equivalent width of
  various emission lines or the 4000\AA break strength, which are less
  sensitive to dust attenuation.

\end{enumerate}

The  second of these  issues is  probably  the most important one. For
example, as shown by Hogg \etal (2004) and Kauffmann \etal (2004), the
colour-magnitude   relation  of   galaxies   is   strongly environment
dependent,   with lower density     environments  showing a  far  more
pronounced blue sequence than high density environments.  Our results,
however, suggest that the colour-magnitude  relation of {\it satellite
  galaxies} has only a  very weak environment dependence.  We believe,
though, that these  results are not inconsistent  with each other, but
merely reflect  that  different environments have  different satellite
fractions: in high density environments, such as clusters of galaxies,
the (vast) majority  of all galaxies are  satellites.  In  low density
environments, however, most galaxies are  central galaxies in low mass
haloes.    In paper~I we  have shown   that central  galaxies  are, on
average, bluer  than satellite galaxies of  the same stellar mass, and
we believe  that  this is  largely   responsible  for the  environment
dependence noticed in the studies mentioned above.
 
The  study whose results  seems most  antithetical to  those presented
here is that of Weinmann \etal (2006a; hereafter W06), who used a SDSS
galaxy  group catalogue very  similar to  that used  here in  order to
study the fractions  of red and blue satellite  (and central) galaxies
as a function of halo mass and absolute $r$-band magnitude (see Hester
2006b for a similar analysis, though they did no split their sample in
centrals and  satellites).  W06 find that at  fixed absolute magnitude
the fractions of red and  blue satellites are strong functions of halo
mass, while at fixed halo mass there is only a very mild dependence on
luminosity.  At  first sight these results are  completely opposite to
the results  presented here\footnote{Although  the results of  W06 are
  based on  $r$-band magnitudes, while those presented  here are based
  on stellar masses, we have verified that our results hold if we were
  to use the $r$-band magnitudes instead.}. However, the following two
reasons show that they are, in fact, consistent with each other.

First  of all,  the  fact  that W06  found  no significant  luminosity
dependence  of the  red fraction  of satellites,  $f_{\rm  red}$, owes
entirely to the  fact that they used a red/blue  dividing line that is
magnitude dependent.   Clearly, for each magnitude bin  one can always
identify  a colour  so that  the satellite  fraction red-ward  of this
colour is independent of luminosity.  Apparently, this is roughly what
happens  when  using  the  magnitude-dependent  cut adopted  by  W06.  
However,  since  the  dependence   of  colour  on  luminosity  is  now
`absorbed' by the magnitude dependence  of the colour-cut, one can not
use the absence  of a luminosity dependence of  $f_{\rm red}$ to argue
that colour does not depend on luminosity.

Secondly, the fact that  W06 found a  significant halo mass dependence
for $f_{\rm red}$ is consistent with our results.  This is most easily
seen from the upper left-hand  corner of Fig.~\ref{fig:his_col}. If we
define  the red satellite fraction,  $f_{\rm  red}$, as the ($1/V_{\rm
  max}$-weighted)  fraction       of      satellites   red-ward     of
$^{0.1}(g-r)=0.7$,  one obtains values  of $0.33$,  $0.47$, and $0.66$
for the  halo  mass bins $[12.0,12.5]$ (red  histogram), $[13.0,13.5]$
(black  histogram)  and $[14.0,14.5]$  (blue histogram), respectively.
Thus, in agreement with W06, we find that $f_{\rm red}$ increases with
increasing halo mass. Upon closer   inspection, it is clear that  this
largely  owes  to  the fact that   the  colour-distributions for these
different  halo mass bins have a  different  skewness.  However, as we
have shown, the first and second moments are very similar. What really
matters  are the actual distribution functions,  and there is no doubt
from Fig.~\ref{fig:his_col} that these depend more strongly on stellar
mass than on halo mass.

We  therefore conclude  that  our results  are  not inconsistent  with
previous studies.  We  do caution, though, that one  has to be careful
not to  overinterpret trends (or the  absence thereof) in  red or blue
fractions. Since  colour is an actual physical  quantity, while $f_{\rm
  red}$ is not  (and is sensitive to a  somewhat arbitrary colour cut),
we believe  that our study presented  here is more  useful for gaining
insight regarding the physical processes at work.

\section{Conclusions}
\label{sec:concl}

Using  a large  SDSS  galaxy  group catalogue,  we  have examined  the
ecology of satellite galaxies. In  particular, we have studied how the
colours  and  concentrations of  satellite  galaxies  depend on  their
stellar mass  and their environment. The latter  is `parameterized' in
terms of the  mass of the host halo and  the halo-centric radius.  Our
main  result  is that  both  the colour  and  the  concentration of  a
satellite  galaxy are  almost completely  determined by  their stellar
mass, with only a very mild  dependence on environment.  Or, to put it
in  different words,  in order  to  predict the  (average) colour  and
concentration  of a satellite  galaxy, all  one needs  to know  is its
stellar mass.  No information  regarding its environment (halo mass or
halo-centric radius) is required.

There is a weak trend  that satellites  at smaller halo-centric  radii
and  in   more  massive   haloes are,  on   average,   redder and more
concentrated. However, we argue  that neither of these are reflections
of a causal    environment  dependence.  Rather,  they owe    to  mass
segregation  (more massive satellites   reside on smaller halo-centric
radii) and to the fact that more massive  satellites on average reside
in more massive haloes. These two trends,  combined with the fact that
more massive satellites are  redder and more concentrated, explain the
weak environment     dependencies evident in    the data;  they almost
completely disappear when using  stellar  mass as a control   variable
(i.e.,  when only considering  satellite galaxies  in  a narrow bin in
stellar mass).

It is interesting  to combine these results with  those of Paper~I, in
which we used the same  SDSS group catalogue to investigate the colour
and concentration differences  between central and satellite galaxies,
in  a  statistical sense.   That  analysis  has  shown that  satellite
galaxies are,  on average, redder and somewhat  more concentrated than
central galaxies of the same  stellar mass.  In addition, it was shown
that the average  magnitude of this difference is  independent of halo
mass, in  excellent agreement with the lack  of environment dependence
presented here.   Combining all  these results, the  following picture
emerges: galaxies  become redder  and somewhat more  concentrated once
they  fall into  a bigger  halo (i.e.,  once they  become  a satellite
galaxy).   This is a  clear manifestation  of environment  dependence. 
However,   there  is  no   indication  that   the  magnitude   of  the
transformation  (or its  timescale) depends  on environment;  a galaxy
undergoes a  transition when it becomes  a satellite, but  it does not
matter whether  it becomes  a satellite of  a small (Milky  Way sized)
halo,  or  of  a  massive  cluster.   In  that  respect  there  is  no
(significant)  environment dependence,  and  this is  the `dearth'  of
environment dependence we are refering to in the title of this paper.

This dearth of environment dependence  puts interesting constraints on
the physical  processes    responsible  for transforming     satellite
galaxies.   As  discussed  at  length in Paper~I,   the  data are most
consistent with a picture in which strangulation, i.e., the removal of
the (hot) gas reservoir of  satellite galaxies, is the main  mechanism
that    operates   on satellite    galaxies,   and  that  causes their
transformation from   the blue to the  red  sequence.   The  fact that
satellites in clusters  have the same  properties as satellites of the
same stellar mass   in galaxy-sized  haloes, strongly  argues  against
mechanisms  that are thought to  operate only in  very massive haloes,
such as ram-pressure  stripping  or galaxy harassment.  Note,  though,
that we are not claiming that these processes do not occur, after all,
clear-cut example of ram-pressure  stripping have been observed (e.g.,
Gavazzi  \etal 1995; Kenney, van  Gorkum \&  Vollmer 2004).  We mainly
argue   that  they are    not the  {\it   dominant}  processes causing
satellites to undergo   a  blue-to-red sequence  transition.   We also
emphasize that we have used the term `ram-pressure stripping' to refer
to a (rapid) stripping  of the {\it   entire} cold gas reservoir  of a
galaxy.   Although we claim that this  can not be   the main cause for
satellite transformations, our   data is perfectly  consistent with  a
picture in which ram-pressure stripping  operates on the cold gas, but
only manages to strip the outer layers of  the gas disk (i.e., the gas
at large galacto-centric radii, where no or very little star formation
occurs). Such a stripping, after all, has basically the same effect as
strangulation (the removal of the  hot gas), as it effectively removes
the  fuel for {\it future}  star formation (Hester 2006a,b).  In fact,
the  observed HI deficiency  of cluster  galaxies (e.g., Warmels 1988;
Cayatte \etal 1990; Bravo-Alfaro \etal 2000; Levy \etal 2007) strongly
suggests that  a mechanism is stripping  some, but not  all, of the HI
gas\footnote{Although  this   is   often  attributed   to ram-pressure
  stripping, viscous stripping (Nulsen 1997) or stripping due to tidal
  forces   could  also play  an    important role.   In  addition, the
  deficiency   may  also simply  be   a  reflection of  the  fact that
  star-formation has used  up a large  fraction of the HI gas, without
  being replenished with new gas since its hot  gas reservoir has been
  removed.}.

Our  results  also argue  against  `pre-processing'  in  groups as  an
important process: as argued above, the only way in which a cluster is
a `special' environment, is that it has a larger fractional population
of satellites than the `field', and that the average satellite mass is
higher than in the field.  It has been suggested, though, that cluster
galaxies are different from their field counterparts because they were
part of an  intermediate mass group prior to  falling into the cluster
(Zabludoff \&  Mulchaey 1998; Zabludoff 2002).  Since  merging is more
effective  in  groups than  in  either  clusters  or the  field,  this
``pre-processing''  could  explain  the morphology-density  relation.  
However,  our  results  show  that  there  is  no  difference  between
satellite galaxies in clusters and satellite galaxies {\it of the same
  stellar  mass}  in  any  other  environment.   This  indicates  that
pre-processing  in   groups  can  not   be  the  dominant   cause  for
differentiating the cluster galaxy population  from that of the field. 
A similar conclusion was recently reached by Berrier \etal (2008), who
used numerical simulations  to show that the vast  majority of cluster
galaxies ($\sim  70$\%) fall into the cluster  potential directly from
the field,  without having  been a satellite  in an  intermediate mass
group environment.

We  conclude  by  postulating,  based  on  our  results,  that  galaxy
properties  depend predominantly  on  stellar mass.   The second  most
important `parameter' for a galaxy is the distinction between centrals
and satellites. When using  these two parameters as control variables,
there is  virtually no  environment dependence left.   This conclusion
has  recently  received support  from  various directions.   Kauffmann
\etal (2004) have shown that at fixed stellar mass, there is nearly no
dependence   of   structural  properties   like   S\'ersic  index   or
concentration  parameter  on  local  galaxy density.   Pasquali  \etal
(2007) have  shown  that  the   trends  of  the  isophotal  shapes  of
elliptical  galaxies  (`disky' vs.  `boxy')  with  environment can  be
completely  explained in  terms of  a  pure stellar  mass dependence.  
Using  the same  group catalogue  as  that used  here, Pasquali  \etal
(2008) have shown that the star formation and AGN activity of galaxies
has a  much stronger  dependence on  stellar mass than  on halo  mass. 
Mouhcine,  Baldry  \&  Bamford   (2007)  find  no  dependence  of  the
relationship  between   galaxy  stellar  mass   and  gas-phase  oxygen
abundance on local galaxy density. Blanton \etal (2005b) and Kauffmann
\etal (2004) find that there is  only a very weak dependence of galaxy
size on local density at fixed luminosity, which, as shown by Weinmann
\etal (2008), is mainly a  reflection of a difference between centrals
and satellites.

\section*{Acknowledgments}

FvdB thanks the  Aspen Center for Physics where part  of this work has
been done,  and is grateful  to Andreas Berlind, Alison  Coil, Charlie
Conroy,  Erin Sheldon,  Jeremy Tinker,  and Risa  Wechsler  for lively
discussions during the Aspen Workshop ``Modelling Galaxy Clustering''.
We  are grateful to  Michael Blanton,  Cheng Li  and Marco  Barden for
their help  with the SDSS,  and to Eric  Bell and Hans-Walter  Rix for
critical remarks.


\end{document}